\documentclass{raa}

\usepackage[T1]{fontenc}
\usepackage{graphicx,times}
\usepackage{xcolor}
\newcommand{\added}[1]{#1}
\newcommand{\revcolor}{}
\usepackage{natbib}
\usepackage{amssymb,amsmath}
\usepackage{booktabs}
\usepackage{multirow}
\usepackage[section]{placeins}
\usepackage[hidelinks]{hyperref}
\bibpunct{(}{)}{;}{a}{}{,}
\date{}
\begin{document}

\title{Late-time distance consistency and model-independent curvature inference with strong lensing, quasars, and galaxy clusters}

\volnopage{ {\bf 20XX} Vol.\ {\bf X} No.\ {\bf XX}, 000--000}
\setcounter{page}{1}

\author{Jian Hu\inst{1,*}, Yi Liu\inst{2}, Jian-Ping Hu\inst{3,4}, Zhongmu Li\inst{1}}

\institute{Institute of Astronomy and Information, Dali University, Dali 671003, China; {\it dg1626002@smail.nju.edu.cn}\\
\and Department of Engineering, Dali University, Dali 671003, People's Republic of China\\
\and Ministry of Education Key Laboratory for Nonequilibrium Synthesis and Modulation of Condensed Matter, School of Physics, Xi'an Jiaotong University, Xianning West Road, Xi'an, China\\
\and Key Laboratory of Modern Astronomy and Astrophysics (Nanjing University), Ministry of Education, Xianlin Road, Nanjing, China\\
\vs \no
{\small Received 20XX Month Day; accepted 20XX Month Day}}

\abstract{We revisit a late-time, model-independent distance-consistency framework that combines strong gravitational lensing systems (SGLS), ultraviolet/X-ray quasars (QSOs), and 38 Sunyaev--Zel'dovich/X-ray galaxy-cluster angular-diameter distances. The analysis focuses on systematics rather than on presenting a competitive curvature measurement. Starting from the 158-system SGLS catalog, we exclude 30 systems that violate the lensing-geometric requirement $D_{A,ls}/D_{A,s}<1$ in the isothermal reference case $\gamma_{\rm lens}=2$, leaving a fixed 128-system SGLS parent sample for cross-matching with 1598 QSOs. A deterministic maximum-cardinality, minimum-mismatch one-to-one assignment yields 100 SGLS--QSO pairs at the nominal $\Delta d/d=5\%$ threshold. We propagate the measured X-ray flux uncertainties and the intrinsic QSO scatter, and, crucially, we include the full $38\times38$ covariance induced by using the same QSO distances in the kernel-smoothed cluster predictions. With $\Omega_K\sim\mathcal{U}[-10,10]$, the joint posterior is $\Omega_K=2.085^{+1.634}_{-1.103}$ (68\%; central 95\% interval $[0.269,6.028]$). We do not interpret the positive shift as evidence for an open Universe. \added{This positive posterior offset is driven by cross-probe calibration tensions, sample-assignment effects, and the cluster anchor; it cannot be attributed to the intrinsic spatial geometry of the Universe.} The SGLS+QSO data alone leave curvature essentially prior dominated, $\Omega_K=0.70^{+6.33}_{-7.12}$, whereas removing only the off-diagonal cluster covariance moves the posterior to $\Omega_K=5.21^{+2.46}_{-2.07}$. Thus the cluster anchor supplies most of the curvature constraining power and its covariance materially changes the posterior location. The prediction-only cluster correlation matrix has a median off-diagonal coefficient $\bar\rho=0.350$ and spectral effective dimension $N_{\rm eff}=2.59$. \added{Shared QSO tracers reduce this prediction component to roughly three effective correlation modes; this redundancy helps explain its covariance sensitivity but is not a literal count of independent constraints in the total cluster likelihood.} Matching-threshold tests over 3--7\% yield 97--103 pairs; the 3, 4, 5, and 7\% posteriors are broadly similar, while the 6\% assignment gives a larger positive excursion. Bayesian evidence favors a free QSO slope over $\gamma_{\rm QSO}=0.6$, with $\ln B_{\rm free/fixed}=4.50\pm0.17$. A 50/50 split at the median source redshift, $z_{s,\rm med}=0.9335$, gives $\gamma_{\rm QSO}=0.351^{+0.057}_{-0.037}$ at low $z_s$ and $0.666^{+0.031}_{-0.038}$ at high $z_s$, \added{indicating redshift-dependent calibration or reconstruction systematics.} \added{The Hubble-constant posterior remains prior dominated and provides no meaningful constraint on $H_0$ or the Hubble tension.} We conclude that the present data are more useful as a diagnostic of cross-probe calibration, covariance, and sample-selection systematics than as a precision measurement of cosmic curvature.
}

\authorrunning{Jian Hu et al.}
\titlerunning{Late-time distance consistency and curvature}
\maketitle

\section{Introduction}\label{sect:intro}

The spatial curvature parameter $\Omega_K$ is a fundamental descriptor of the large-scale geometry of the Universe. We adopt the standard convention in which $\Omega_K>0$ corresponds to an open geometry, $\Omega_K=0$ to a spatially flat geometry, and $\Omega_K<0$ to a closed geometry. Within non-flat $\Lambda$CDM, combinations of CMB lensing and low-redshift BAO data are consistent with a nearly flat Universe, for example $\Omega_K=0.0007\pm0.0019$ in the Planck analysis \citep{planck20}. Planck temperature and polarization data without the same low-redshift geometric information can exhibit a preference for negative $\Omega_K$ (closed geometry), illustrating the model and data-combination dependence of curvature inference \citep{di20}. \added{Earlier Planck results and the BOSS DR12 analysis likewise demonstrate the role of CMB and BAO combinations in geometric constraints \citep{planck16,alam17}. Late-time supernova, BAO, and $H(z)$ analyses provide complementary constraints within specified cosmological models \citep{para19}. Phenomenological averaged-cosmology studies further illustrate why the assumptions connecting distances to spatial curvature must be stated explicitly \citep{san17,cole19}.}

A separate line of work seeks curvature information without specifying a dark-energy model. The Clarkson consistency relation combines a distance observable with $H(z)$ and its derivative \citep{cla07,cla08}, while the distance-sum-rule approach relates dimensionless comoving distances between observer, lens, and source \citep{ber06,ras15}. The latter is particularly attractive for strong-lensing applications because the lensing observable directly provides a distance ratio. \citet{ras15} combined strong lenses with standard-candle distances and obtained a broad, model-independent curvature constraint.

\added{Subsequent distance-sum-rule analyses examined the sensitivity to lens populations and mass models \citep{xia17,li18,wang20}, while strong-lensing time delays combined with supernovae have enabled joint geometric inference of $H_0$ and curvature \citep{col19}. A complementary construction compares proper and transverse comoving distances using $H(z)$ and angular-diameter distances \citep{yuw16}. Supernova compilations such as Union2.1 and JLA \citep{suz12,bet14}, and model-independent calibration of their effective magnitudes \citep{hu23b}, provide additional context for constructing calibrated late-time distance networks.}

Quasar Hubble diagrams based on the nonlinear X-ray--UV luminosity relation have subsequently been used in several cosmological and model-independent distance studies \added{\citep{ris19,lusso20,liu20,lenart23,ocolgain22}}. Their advantage is their redshift reach; their limitation is that the slope and normalization of the luminosity relation are strongly coupled to the inferred absolute distance scale. Compact radio quasars provide a complementary approach because their angular sizes can be used as direct angular-diameter-distance tracers, thereby avoiding the same X-ray--UV calibration degeneracy \citep{cao17}. We therefore do not claim that UV/X-ray QSOs are uniquely suited to curvature measurements. Our purpose is instead to use them as one member of an explicitly cross-calibrated late-time distance network and to quantify where that construction fails.

The second ingredient in the present analysis is a set of 38 angular-diameter distances to galaxy clusters from the joint SZ/X-ray analysis of \citet{bonamente06}. These measurements are individually noisy, with a median fractional observational uncertainty of about 19\% in the sample used here, but they carry absolute-distance information that is independent of the QSO flux normalization. Their value in this framework is therefore not per-object precision; it is their ability to anchor an otherwise weakly determined absolute distance scale. The same feature also makes the cluster term a sensitive diagnostic of cross-probe inconsistency.

Our analysis has three methodological components that are central to the interpretation. First, the SGLS--QSO sample is constructed with a deterministic global assignment and is tested over 3--7\% matching thresholds. Second, we propagate the covariance among all 38 kernel-smoothed cluster predictions rather than treating them as independent. Third, SGLS+QSO-only and Bayesian-evidence control runs identify which conclusions are supplied by the cluster anchor and which remain prior or calibration dependent.

The resulting curvature posterior is shifted to positive values, but its location changes strongly under covariance and matching choices. We therefore treat $\Omega_K$ as a \emph{diagnostic consistency parameter} in this data combination, not as a competitive curvature measurement.

\section{Data and sample construction}\label{sect:data}

\subsection{Quasar sample}\label{sect:qso}

We use the 1598-object QSO catalog of \citet{ris19}, spanning approximately $0.04\lesssim z\lesssim5.1$. The X-ray--UV luminosity relation is written
\begin{equation}
\log_{10}L_X=\gamma_{\rm QSO}\log_{10}L_{\rm UV}+\beta,
\end{equation}
and, in flux form,
\begin{equation}\label{eq:fluxrel}
\log_{10}F_X=\beta'+\gamma_{\rm QSO}\log_{10}F_{\rm UV}+2(\gamma_{\rm QSO}-1)\log_{10}D_L.
\end{equation}
The internal code normalization $\beta_{\rm code}$ differs from the luminosity-space $\beta$ by the usual flux and distance-unit conversions. In the inference below $\beta_{\rm code}$ is not externally calibrated.

The parent table provides an uncertainty for $\log F_X$ but not a homogeneous UV-flux uncertainty column for the matched sample. We therefore propagate the measured X-ray uncertainty together with an intrinsic dispersion parameter $\delta$ and do not invent an unreported UV error term.

\subsection{Strong-lensing parent sample}\label{sect:sgls}

The parent strong-lensing catalog contains 158 systems compiled from the samples used by \citet{leme18}, \citet{cao15}, and \citet{shu17}. Before any SGLS--QSO cross-matching, we impose the lensing-geometric admissibility requirement $0<R_{\rm lens}<1$ in the isothermal reference case $\gamma_{\rm lens}=2$, where
\begin{equation}\label{eq:rlens}
R_{\rm lens}\equiv\frac{D_{A,ls}}{D_{A,s}}=
\frac{\theta_E c^2}{4\pi\sigma_{\rm ap}^2 f(\gamma_{\rm lens})}
\left(\frac{\theta_{\rm ap}}{\theta_E}\right)^{2-\gamma_{\rm lens}},
\end{equation}
with
\begin{equation}
f(\gamma)= -\frac{(5-2\gamma)(1-\gamma)}{\sqrt{\pi}(3-\gamma)}
\frac{\Gamma(\gamma-1)}{\Gamma(\gamma-3/2)}
\left[\frac{\Gamma(\gamma/2-1/2)}{\Gamma(\gamma/2)}\right]^2.
\end{equation}
Systems with $R_{\rm lens}\ge1$, equivalently $D_{A,ls}\ge D_{A,s}$ under this reference definition, are excluded before catalog matching. This single geometric cut removes 30 systems and leaves 128 physically admissible SGLS systems. These 128 systems form the fixed SGLS parent sample supplied to the matching algorithm; no additional SGLS preselection is performed during threshold matching. The parent-sample audit is summarized in Appendix~\ref{app:selection}.

\added{The lens-density slope is retained as a nuisance parameter because possible redshift dependence and calibration sensitivity of the power-law lens model have been investigated using both cosmological-model-dependent and model-independent approaches \citep{cui17,hol17,hu23a}. The isothermal reference used for the initial sample cut does not fix the slope in the subsequent likelihood.}

\subsection{Deterministic SGLS--QSO matching}\label{sect:matching}

The fixed 128-system SGLS parent sample and the 1598-QSO catalog are matched in a dimensionless comoving-distance coordinate calculated solely for catalog association in a flat reference model with $\Omega_m=0.31$. For each SGLS target redshift ($z_l$ and $z_s$), a QSO is eligible for the corresponding lens-side or source-side role when
\begin{equation}\label{eq:matching_tol}
\frac{|d_{\rm QSO}-d_{\rm target}|}{d_{\rm target}}\le \epsilon.
\end{equation}
The reference cosmology is not used in the subsequent likelihood. A distance-based matched-pair criterion was also used in our recent model-independent CDDR analysis of the Bonamente cluster compilation and Pantheon+ SNe Ia, including explicit 3\% and 5\% tolerance tests \citep{hu26cddr}.

The matching implementation follows the deterministic logic of \textsc{CosmoMatcher} \citep{cosmomatcher}. For computational efficiency, the complete candidate universe is generated once at the largest tested threshold, $\epsilon=0.07$, and is then filtered to $\epsilon=0.03,0.04,0.05,0.06,$ and 0.07. For a candidate triplet consisting of one SGLS, one lens-side QSO, and one source-side QSO, the required tolerance is
\begin{equation}
\epsilon_{\rm req}=\max\!\left(\frac{|d_{q,l}-d_l|}{d_l},\frac{|d_{q,s}-d_s|}{d_s}\right),
\end{equation}
and the mismatch cost is
\begin{equation}
C=\left(\frac{|d_{q,l}-d_l|}{d_l}\right)^2+\left(\frac{|d_{q,s}-d_s|}{d_s}\right)^2.
\end{equation}
The same QSO cannot occupy both roles in a single triplet. We then solve a lexicographic mixed-integer linear program: the number of matched SGLS systems is maximized first, and among all maximum-cardinality solutions the total mismatch cost is minimized. Each SGLS is used at most once, and each QSO is used at most once globally across both lens-side and source-side roles.

At the nominal $\epsilon=0.05$ threshold, this procedure returns 100 independent SGLS--QSO pairs using 200 distinct QSOs, with SGLS lens redshifts $0.0625\le z_l\le1.004$ and source redshifts $0.324\le z_s\le3.595$. The 3, 4, 5, 6, and 7\% thresholds yield 97, 100, 100, 101, and 103 matched systems, respectively. No QSO-derived $d_l/d_s$ post-selection depending on $\gamma_{\rm QSO}$ is applied after the global assignment. The archived implementation is available at \url{https://doi.org/10.5281/zenodo.19695657}, with the development repository at \url{https://github.com/HUJIAN0000/CosmoMatcher}.

\subsection{Galaxy-cluster angular-diameter distances}\label{sect:clusters}

We use the 38 SZ/X-ray angular-diameter distances compiled by \citet{bonamente06}, covering $0.14\le z_c\le0.89$. The quoted asymmetric uncertainties are symmetrized as the mean of the upper and lower errors. The median fractional observational uncertainty is 19.1\%. Cluster morphology and temperature structure remain important systematics; this limitation is discussed in Sec.~\ref{sect:discussion}.

\section{Method}\label{sect:method}

\subsection{Distance-sum-rule likelihood}\label{sect:sumrule}

For the sign convention used here, $k\equiv-\Omega_K$, and the distance sum rule gives
\begin{equation}\label{eq:sumrule}
R_{\rm QSO}=\sqrt{1+\Omega_K d_l^2}-\frac{d_l}{d_s}\sqrt{1+\Omega_K d_s^2}.
\end{equation}
The dimensionless distance reconstructed from the QSO flux relation is
\begin{equation}\label{eq:qso_d}
d(z)=\frac{H_0}{c}\frac{D_L(z)}{1+z},
\end{equation}
with $D_L$ obtained from Eq.~(\ref{eq:fluxrel}). Because the matched QSO is used as a distance tracer for the target SGLS redshift, the distance conversion is evaluated at the corresponding SGLS $z_l$ or $z_s$; the actual QSO redshift enters only the catalog-matching step.

The uncertainty in the logarithmic luminosity distance is
\begin{equation}\label{eq:qsoerr}
\sigma_{\log D_L}=\frac{\sqrt{\delta^2+\sigma_{\log F_X}^2}}{|2(\gamma_{\rm QSO}-1)|},
\end{equation}
so that $\sigma_d/d=\ln(10)\,\sigma_{\log D_L}$. For Eq.~(\ref{eq:sumrule}) we use the corrected derivatives
\begin{align}
\frac{\partial R}{\partial d_l}&=\frac{\Omega_K d_l}{\sqrt{1+\Omega_K d_l^2}}-\frac{\sqrt{1+\Omega_K d_s^2}}{d_s},\\
\frac{\partial R}{\partial d_s}&=\frac{d_l}{d_s^2\sqrt{1+\Omega_K d_s^2}},
\end{align}
which are propagated together with the observational uncertainty in Eq.~(\ref{eq:rlens}). The SGLS+QSO log-likelihood is the Gaussian likelihood of $R_{\rm lens}-R_{\rm QSO}$ including its normalization term.

\subsection{Full covariance of the cluster anchor}\label{sect:covmethod}

At each cluster redshift $z_c^i$ we estimate the QSO-implied dimensionless distance with a Gaussian Nadaraya--Watson kernel,
\begin{equation}
\hat d_i=\sum_j W_{ij}d_j,\qquad
W_{ij}=\frac{\exp[-(z_j-z_c^i)^2/(2h^2)]}{\sum_m\exp[-(z_m-z_c^i)^2/(2h^2)]},
\end{equation}
with $h=0.10$ and with the sum over both lens-side and source-side matched QSO distance tracers. \added{The bandwidth and the tracer set determine $\mathbf W$; therefore a bandwidth or redshift-bin change modifies both the predicted distances and their propagated covariance in Eq.~(\ref{eq:cpred}).} The predicted cluster angular-diameter distance is
\begin{equation}
D_{A,i}^{\rm pred}=\frac{c}{H_0}\frac{\hat d_i}{1+z_c^i}.
\end{equation}

Because the same QSO measurements contribute to several cluster predictions, the prediction errors are correlated. Let $S={\rm diag}[c/H_0/(1+z_c^i)]$, and let $\boldsymbol{\Sigma}_d={\rm diag}(\sigma_{d_j}^2)$. The induced prediction covariance is
\begin{equation}\label{eq:cpred}
\mathbf C_{\rm pred}=\mathbf S\,\mathbf W\,\boldsymbol{\Sigma}_d\,\mathbf W^{\rm T}\mathbf S,
\end{equation}
and the total cluster covariance is
\begin{equation}
\mathbf C=\mathbf C_{\rm obs}+\mathbf C_{\rm pred},
\end{equation}
where $\mathbf C_{\rm obs}$ is diagonal with the published cluster variances. The cluster likelihood is
\begin{equation}\label{eq:clusterll}
\ln\mathcal L_{\rm cl}=-\frac12\left[\mathbf r^{\rm T}\mathbf C^{-1}\mathbf r+\ln\det\mathbf C+N_{\rm cl}\ln(2\pi)\right],
\end{equation}
with $\mathbf r=\mathbf D_A^{\rm obs}-\mathbf D_A^{\rm pred}$ and $N_{\rm cl}=38$. This \added{full cluster covariance} is used in the nominal likelihood; the diagonal approximation is retained only as a dedicated control test.

\subsection{Priors and sampling}\label{sect:priors}

We sample the six-parameter vector
\begin{equation}
\boldsymbol\theta=(\gamma_{\rm lens},\gamma_{\rm QSO},H_0,\delta,\Omega_K,\beta_{\rm code}).
\end{equation}
The priors are
\begin{align*}
\gamma_{\rm lens}&\sim\mathcal U[1.5,2.5], &
\gamma_{\rm QSO}&\sim\mathcal U[0.3,1.0],\\
\delta&\sim\mathcal U[0.001,1.0], &
\Omega_K&\sim\mathcal U[-10,10],\\
\beta_{\rm code}&\sim\mathcal U[-30,-5], &
H_0&\sim\mathcal N(70,5)\ \mathrm{km\,s^{-1}\,Mpc^{-1}},
\end{align*}
with $40<H_0<100$ imposed as a broad numerical box. The $H_0$ prior is informative for this data combination, and we do not interpret the resulting $H_0$ posterior as an independent Hubble-constant measurement.

The baseline posterior is sampled with \texttt{emcee} \citep{emcee13} using 60 walkers, 2000 burn-in steps, and 8000 production steps. For the nominal 5\% run \added{using the full cluster covariance}, the acceptance fraction is 0.503, the maximum integrated autocorrelation time is approximately 75 steps, and the production length is about 107 autocorrelation times.

\section{Results and systematics}\label{sect:results}

\subsection{Nominal 5\% result \added{with full cluster covariance}}\label{sect:baseline}

Table~\ref{tab:baseline} gives the marginalized constraints for the deterministic 5\% / 100-pair sample. The curvature posterior is
\begin{equation}\label{eq:okbaseline}
\Omega_K=2.085^{+1.634}_{-1.103}\quad(68\%),
\end{equation}
with a central 95\% interval $[0.269,6.028]$. Although zero lies just outside this formal central 95\% interval, the tests below show that the posterior location is sensitive to the cluster covariance and to the global matching assignment. We therefore do not interpret Eq.~(\ref{eq:okbaseline}) as evidence for non-zero curvature.

\begin{table}[!htbp]
\caption{Marginalized constraints for the nominal 5\% / 100-pair analysis \added{using the full cluster covariance}.}
\label{tab:baseline}
\centering
\begin{tabular}{lcc}
\toprule
Parameter & 68\% interval & central 95\% interval\\
\midrule
$\gamma_{\rm lens}$ & $1.974^{+0.050}_{-0.046}$ & $[1.889,2.072]$\\
$\gamma_{\rm QSO}$ & $0.442^{+0.049}_{-0.055}$ & $[0.334,0.534]$\\
$H_0$ & $69.30^{+4.99}_{-5.00}$ & $[59.56,79.25]$\\
$\delta$ & $0.275^{+0.030}_{-0.026}$ & $[0.228,0.340]$\\
$\Omega_K$ & $2.085^{+1.634}_{-1.103}$ & $[0.269,6.028]$\\
$\beta_{\rm code}$ & $-15.104^{+0.991}_{-1.117}$ & $[-17.315,-13.248]$\\
\bottomrule
\end{tabular}
\end{table}

The $H_0$ posterior is almost centered on the imposed $\mathcal N(70,5)$ prior and has nearly the same width; this confirms that $H_0$ is prior dominated. The lens-density slope remains close to the isothermal value. \added{No meaningful constraint on the Hubble constant or the Hubble tension problem can be derived from this analysis. In particular, this posterior cannot discriminate between the CMB-based inference and the local distance-ladder measurement \citep{planck20,rie22}.}

\subsection{What the cluster anchor contributes}\label{sect:anchor}

A decisive control is to remove the cluster likelihood while keeping the same 100 SGLS--QSO pairs and the same priors. The SGLS+QSO-only result is
\begin{align}
\gamma_{\rm QSO}&=0.346^{+0.043}_{-0.031},\\
\Omega_K&=0.70^{+6.33}_{-7.12},\\
\beta_{\rm code}&=-23.65^{+4.37}_{-4.31}.
\end{align}
The central 95\% interval for curvature, $[-9.42,9.55]$, spans almost the entire adopted prior. Thus SGLS+QSO alone do not provide a meaningful curvature measurement in this implementation. The cluster distances supply most of the constraining power by anchoring the absolute distance scale and compressing the allowed $(\gamma_{\rm QSO},\beta_{\rm code},\Omega_K)$ volume.

\begin{figure}[!htbp]
\centering
\includegraphics[width=0.95\columnwidth]{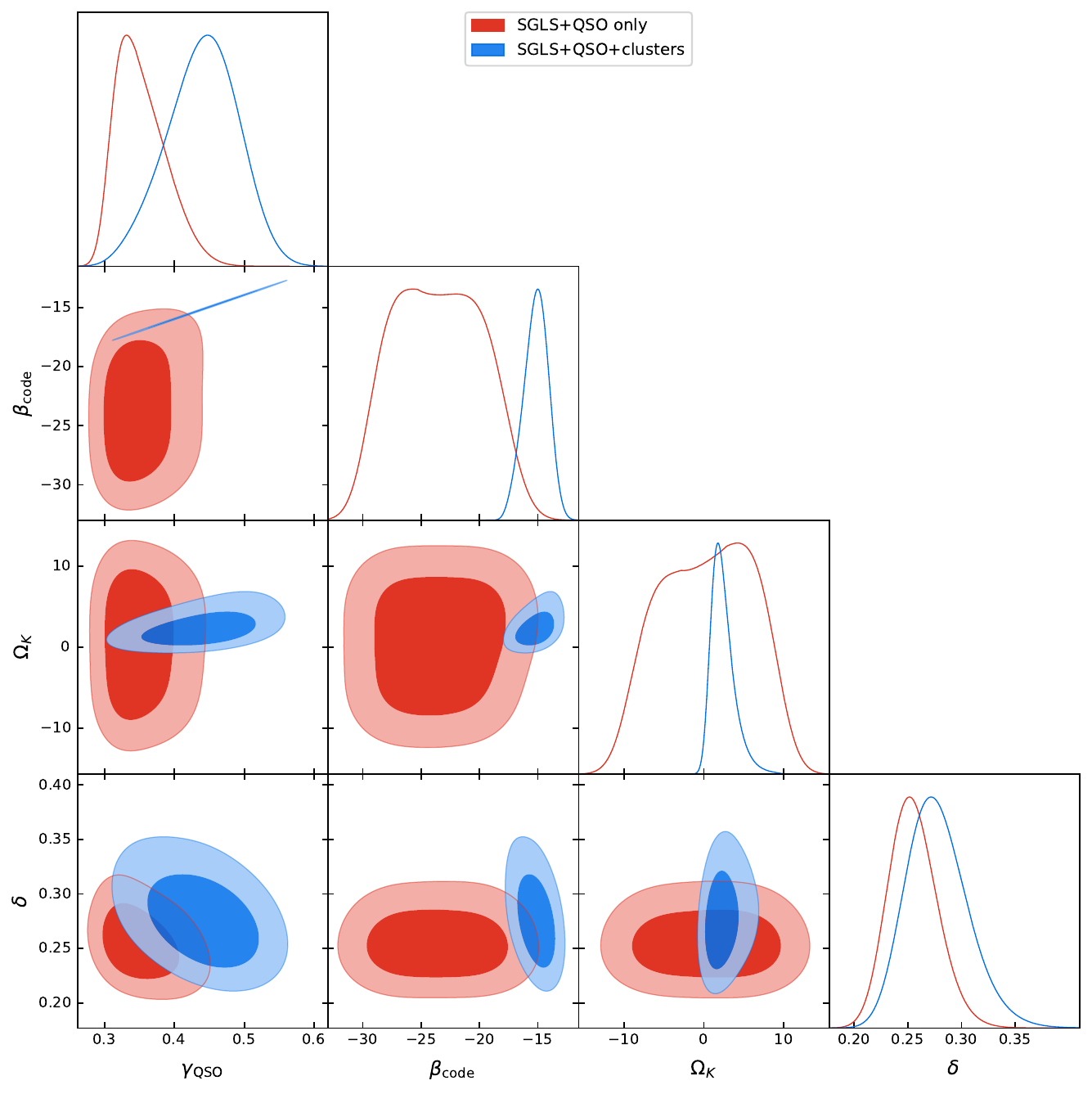}
\caption{Comparison of the SGLS+QSO-only posterior and the joint SGLS+QSO+cluster posterior for $(\gamma_{\rm QSO},\beta_{\rm code},\Omega_K,\delta)$. The cluster anchor substantially narrows the allowed absolute-distance scale, while a tight $\gamma_{\rm QSO}$--$\beta_{\rm code}$ correlation remains in the joint posterior.}
\label{fig:overlay}
\end{figure}

The joint-chain correlation coefficient between $\gamma_{\rm QSO}$ and $\beta_{\rm code}$ is 0.9996. This strong residual ridge is why the curvature posterior should be interpreted together with the QSO calibration parameters rather than in isolation.

\subsection{Effect of the full cluster covariance}\label{sect:covresults}

The prediction-only cluster correlation matrix has median off-diagonal coefficient
\begin{equation}
\bar\rho_{\rm pred}=0.350,
\end{equation}
and a spectral effective dimension
\begin{equation}
N_{\rm eff}=\frac{(\sum_a\lambda_a)^2}{\sum_a\lambda_a^2}=2.59,
\end{equation}
where $\lambda_a$ are the eigenvalues of the prediction correlation matrix. \added{Nearby cluster predictions use overlapping sets of QSO distance tracers, so their prediction errors are strongly correlated. In the spectral sense, $N_{\rm eff}=2.59$ corresponds to roughly three effective anchor modes in the QSO-prediction component. This is a redundancy diagnostic, not a literal count of independent cosmological constraints in the total covariance $\mathbf C_{\rm obs}+\mathbf C_{\rm pred}$. Neglecting the off-diagonal terms treats shared prediction errors as independent and changes the relative weighting of residual modes. In a joint calibration problem this can alter the posterior location as well as its width; the control comparison below demonstrates the size of that effect for the present data.} Observational cluster errors dilute the median off-diagonal correlation of the total covariance to about 0.060, but the shared-QSO component is strongly correlated.

\begin{figure}[!htbp]
\centering
\includegraphics[width=0.78\columnwidth]{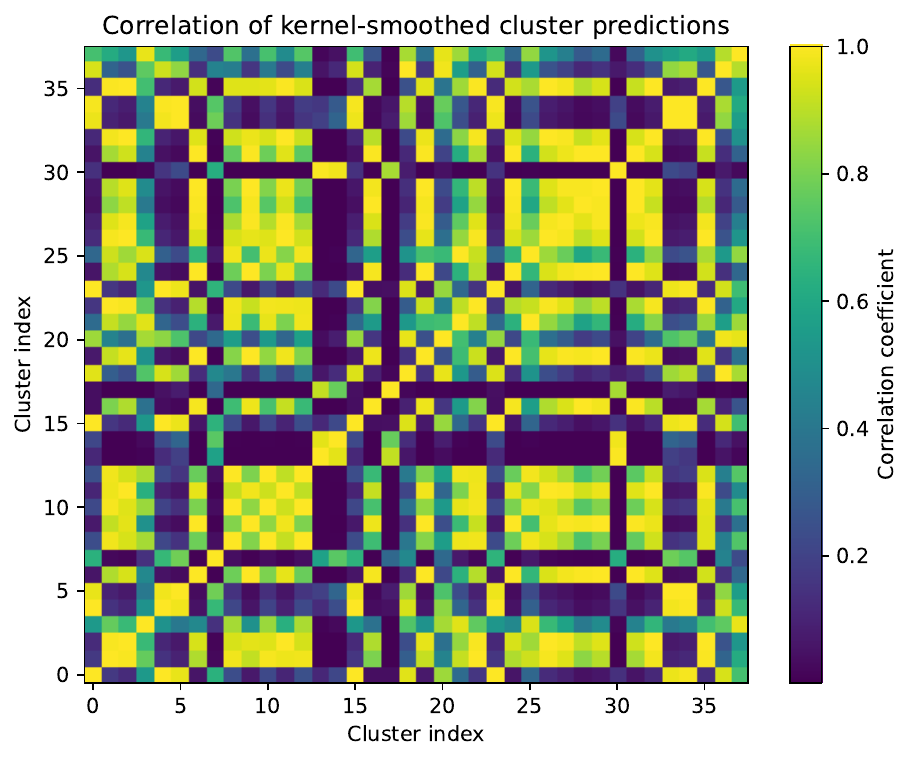}
\caption{Prediction-only correlation matrix of the 38 kernel-smoothed cluster distances evaluated at the nominal joint-posterior median. The broad correlated blocks arise because nearby clusters share many of the same QSO distance tracers.}
\label{fig:corr}
\end{figure}

To isolate the effect of those correlations, we repeat the analysis after setting only the off-diagonal elements of $\mathbf C$ to zero. The result is shown in Table~\ref{tab:covcontrol}. The curvature median moves by $\Delta\Omega_K=-3.13$ when the \added{full cluster covariance} is restored. The 95\% diagonal-control interval $[1.63,9.49]$ remains inside the adopted $[-10,10]$ prior. The key point is not that the \added{full cluster covariance} simply broadens an error bar; rather, it materially changes the location and shape of the joint posterior.

\begin{table}[!htbp]
\caption{\added{Full cluster covariance versus its diagonal approximation.}}
\label{tab:covcontrol}
\centering
\begin{tabular}{lcc}
\toprule
Parameter & \added{Diagonal cluster covariance} & \added{Full cluster covariance}\\
\midrule
$\gamma_{\rm lens}$ & $1.963^{+0.052}_{-0.047}$ & $1.974^{+0.050}_{-0.046}$\\
$\gamma_{\rm QSO}$ & $0.540^{+0.034}_{-0.039}$ & $0.442^{+0.049}_{-0.055}$\\
$\delta$ & $0.293^{+0.031}_{-0.027}$ & $0.275^{+0.030}_{-0.026}$\\
$\Omega_K$ & $5.213^{+2.458}_{-2.067}$ & $2.085^{+1.634}_{-1.103}$\\
$\beta_{\rm code}$ & $-13.138^{+0.699}_{-0.798}$ & $-15.104^{+0.991}_{-1.117}$\\
\bottomrule
\end{tabular}
\end{table}

\subsection{Sensitivity to the matching threshold}\label{sect:threshold}

We repeat the full deterministic assignment and the MCMC \added{using the full cluster covariance} for $\Delta d/d=3,4,5,6,$ and 7\%. Table~\ref{tab:threshold} summarizes the results. The number of matched systems changes only from 97 to 103 even though the number of eligible assignment triplets increases by more than a factor of five. The 3, 4, 5, and 7\% curvature posteriors have similar medians near $2$--$2.5$. The 6\% sample yields a larger positive excursion, $\Omega_K=4.28^{+2.09}_{-1.59}$. \added{The large positive shift of $\Omega_K$ at the 6\% matching threshold is not caused simply by adding one additional matched SGLS--QSO system. Instead, changing the matching threshold modifies the global one-to-one optimization solution and leads to reassignment of multiple QSO--lens pairings, thereby changing the cross-probe distance configuration entering the likelihood. The resulting shift should therefore be interpreted as a sample-assignment systematic associated with the global matching procedure.}

\begin{table}[!htbp]
\caption{Matching-threshold sensitivity. All rows use the full cluster covariance and $\Omega_K\sim\mathcal U[-10,10]$.}
\label{tab:threshold}
\centering
\begin{tabular}{cccc}
\toprule
$\Delta d/d$ & Candidate triplets & Matched pairs & $\Omega_K$ (68\%)\\
\midrule
3\% & 118954 & 97 & $2.463^{+1.818}_{-1.264}$\\
4\% & 208621 & 100 & $2.267^{+1.594}_{-1.091}$\\
5\% & 325131 & 100 & $2.085^{+1.634}_{-1.103}$\\
6\% & 473069 & 101 & $4.281^{+2.093}_{-1.587}$\\
7\% & 638073 & 103 & $2.275^{+1.496}_{-1.048}$\\
\bottomrule
\end{tabular}
\end{table}

\begin{figure}[!htbp]
\centering
\includegraphics[width=0.88\columnwidth]{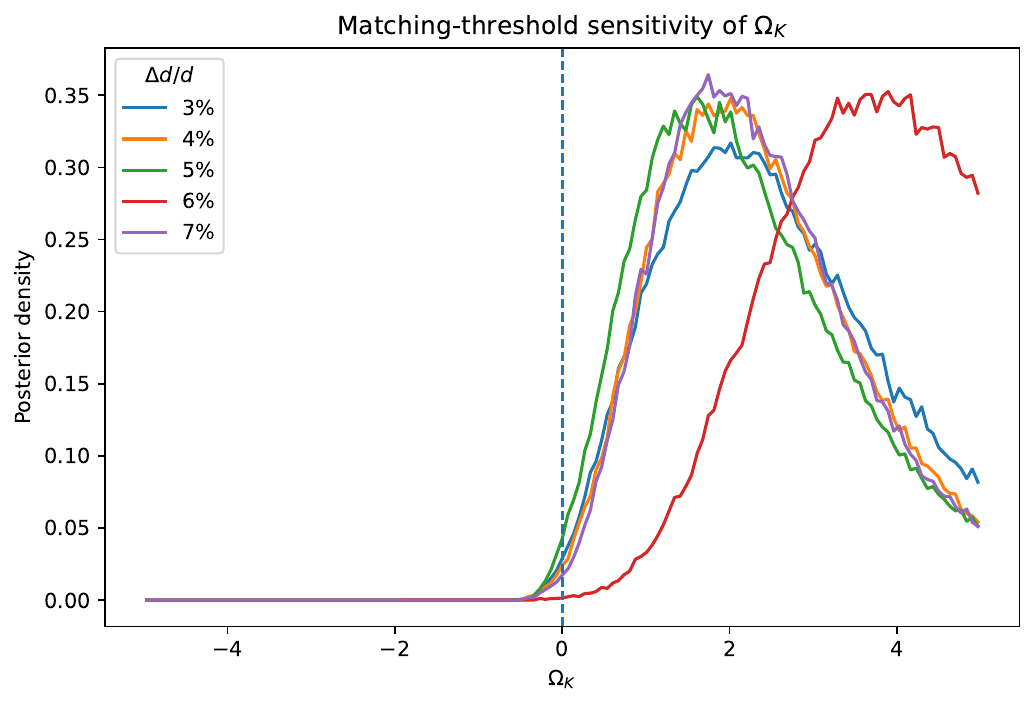}
\caption{Curvature posteriors obtained after reconstructing the global SGLS--QSO assignment at five matching thresholds. The 6\% case is visibly displaced relative to the other four thresholds, demonstrating a non-negligible assignment systematic.}
\label{fig:threshold}
\end{figure}

\subsection{Bayesian evidence for the QSO slope}\label{sect:bayes}

The nominal joint fit prefers $\gamma_{\rm QSO}\simeq0.44$, below the canonical value near 0.6. To quantify whether the data prefer keeping the slope free, we compute the nested-sampling evidence for three models: a free slope with $\gamma_{\rm QSO}\in[0.3,1.0]$, a narrow slope prior $[0.55,0.65]$, and a model fixed at $\gamma_{\rm QSO}=0.6$. All other priors and the likelihood \added{using the full cluster covariance} are identical.

\begin{table}[!htbp]
\caption{Nested-sampling evidence for the QSO-slope models.}
\label{tab:bayes}
\centering
\begin{tabular}{lcc}
\toprule
Model & $\log Z$ & Relative evidence\\
\midrule
Free $\gamma_{\rm QSO}$ & $-362.777\pm0.126$ & reference\\
$0.55<\gamma_{\rm QSO}<0.65$ & $-365.164\pm0.122$ & $\ln B_{\rm free/narrow}=2.387\pm0.175$\\
$\gamma_{\rm QSO}=0.6$ & $-367.278\pm0.118$ & $\ln B_{\rm free/fixed}=4.501\pm0.173$\\
\bottomrule
\end{tabular}
\end{table}

The Bayes factor corresponding to the free-versus-fixed comparison is about $B\simeq90$. This result is specific to the present matched catalog and likelihood; it should not be interpreted as a universal rejection of the literature QSO slope. Instead it demonstrates that externally fixing $\gamma_{\rm QSO}=0.6$ is disfavored within this particular cross-probe calibration problem.

\subsection{Cluster--QSO distance consistency}\label{sect:residuals}

At the nominal posterior median the fractional cluster residuals
\begin{equation}
\frac{D_A^{\rm obs}-D_A^{\rm pred}}{D_A^{\rm pred}}
\end{equation}
have a median of $-9.5\%$ and a standard deviation of 36.5\%. The sign differs from the earlier diagonal-likelihood analysis because both the matched catalog and covariance treatment have changed. The scatter remains substantially larger than the median cluster observational error (19.1\%), emphasizing that the cluster anchor is a noisy cross-calibration rather than a high-precision standard ruler in this application.

\begin{figure}[!htbp]
\centering
\includegraphics[width=0.88\columnwidth]{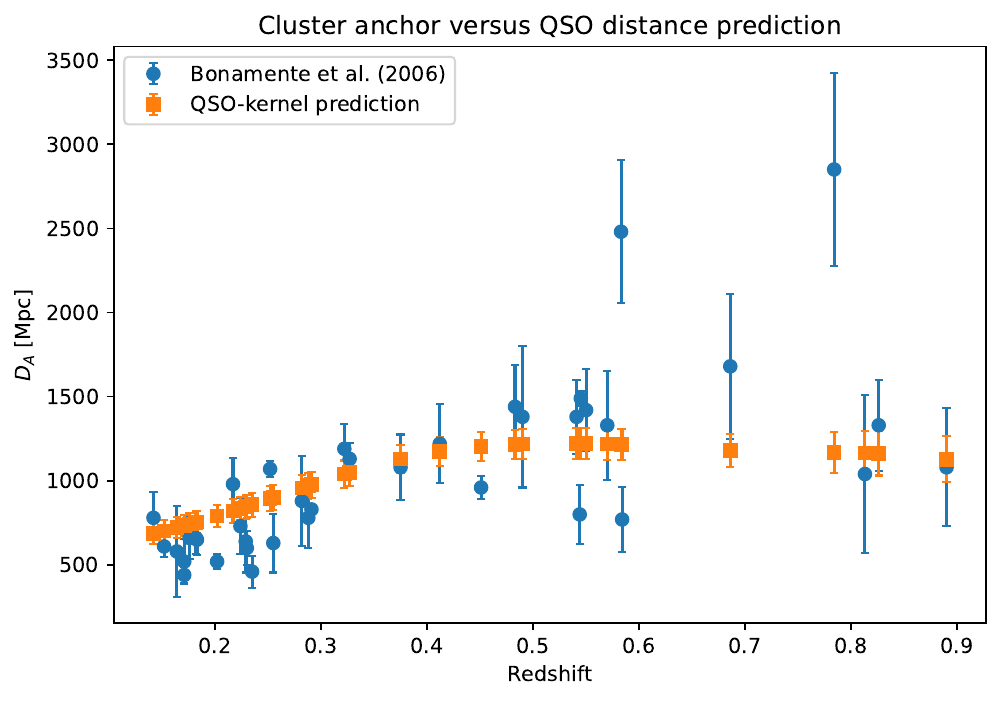}
\caption{Bonamente et al. (2006) cluster angular-diameter distances compared with the QSO-kernel predictions evaluated at the nominal joint-posterior median. The QSO error bars shown here are the propagated kernel-prediction standard deviations before addition of the cluster measurement error.}
\label{fig:clusterconsistency}
\end{figure}

\section{Discussion}\label{sect:discussion}

\subsection{What is and is not learned about curvature}

The current analysis does not yield a robust late-time detection of non-zero curvature. Formally, the 5\% posterior \added{using the full cluster covariance} is shifted to positive $\Omega_K$, but three observations prevent a curvature-detection interpretation. First, SGLS+QSO alone are essentially uninformative over $\Omega_K\in[-10,10]$. Second, omitting the off-diagonal cluster covariance moves the median from 2.09 to 5.21. Third, the global matching assignment produces a noticeable 6\% threshold excursion. The curvature parameter is therefore absorbing part of the residual mismatch among the lensing ratios, the QSO absolute-distance calibration, and the cluster anchor.

The nominal 68\% half-width is approximately $\sigma_{\Omega_K}\simeq1.37$, about $7\times10^2$ times broader than the $\sim0.0019$ Planck+BAO uncertainty quoted above. This is why the present result should not be positioned as a competitive curvature measurement.

\subsection{Rationale for retaining the SZ/X-ray cluster anchor}

The Bonamente sample has large per-cluster uncertainties and a low prediction-only effective dimension after kernel smoothing. Nevertheless, it serves a specific diagnostic function that SGLS+QSO alone cannot provide: an absolute distance anchor that is independent of the QSO flux normalization. The \added{SGLS+QSO-only} control demonstrates this directly. The price is sensitivity to cluster geometry and to the induced covariance, so the cluster term should be viewed as a stress test of cross-probe consistency.

Modern compact-radio-QSO standard rulers are attractive alternatives because they provide angular-diameter-distance information more directly \citep{cao17}. Replacing or augmenting the UV/X-ray QSO Hubble diagram with such probes would reduce reliance on the $(\gamma_{\rm QSO},\beta)$ calibration ridge and is a natural extension of the present framework.

\subsection{Redshift-binned consistency of the QSO calibration}\label{sect:redshift_split}

To test for redshift-dependent behavior of the QSO luminosity relation, we divide the nominal 100-pair sample at the median SGLS source redshift, $z_{s,\rm med}=0.9335$, obtaining 50 low-$z_s$ and 50 high-$z_s$ systems. \added{For the fiducial split, each subsample is re-fit with the same parameter priors, bandwidth $h=0.10$, X-ray flux-error propagation, and full cluster-covariance prescription as the baseline analysis. The two bins contain different QSO tracer sets; using the same prescription does not mean using an identical numerical prediction-covariance matrix.} This is a system-level redshift-consistency test: each lens system contains both a lens-side and a source-side QSO distance tracer, and both redshift bins use the same cluster catalog. The two fits should therefore not be treated as statistically independent measurements of intrinsic QSO evolution. \added{We additionally tested the dependence of the redshift-split result on the fixed-bandwidth Nadaraya--Watson smoothing prescription by repeating the analysis with kernel bandwidths $h=0.05$, $0.10$, $0.15$, and $0.20$ (Table~\ref{tab:bandwidth}). The inferred $\gamma_{\rm QSO}$ and $\Omega_K$ values vary appreciably with bandwidth, indicating sensitivity to the smoothing prescription. Because the kernel-smoothed QSO distances enter the cluster predictions, any redshift-dependent smoothing distortion can propagate into both $\gamma_{\rm QSO}$ and $\Omega_K$. The low--high redshift separation persists over the tested range, showing that it is not specific to the fiducial choice $h=0.10$. However, all four tests use the same fixed-bandwidth Nadaraya--Watson estimator. They do not rule out shared redshift-dependent biases, including boundary effects and sensitivity to the tracer redshift distribution \citep{fan92}. The split is therefore a calibration/reconstruction consistency diagnostic, not a separation of smoothing bias from intrinsic QSO evolution. A comparison with local-linear or adaptive-bandwidth reconstruction would be needed to investigate estimator dependence further.}

\begin{table}[!htbp]
\revcolor
\caption{\added{Kernel-bandwidth sensitivity for the fixed low- and high-$z_s$ subsets of the nominal 100-pair sample (50 systems in each bin). Entries are posterior medians with 68\% credible intervals. The same matching sample, priors, and full cluster-covariance prescription as the fiducial redshift-split analysis are adopted, while each bandwidth case is evaluated with an independent MCMC realization. The fiducial $h=0.10$ redshift-split results are summarized in Table~\ref{tab:redshift}.}}
\label{tab:bandwidth}
\centering
\begin{tabular}{cccc}
\hline
$h$ & bin & $\gamma_{\rm QSO}$ & $\Omega_K$\\
\hline
0.05 & low-$z_s$ & $0.338^{+0.046}_{-0.027}$ & $4.686^{+2.702}_{-2.215}$ \\
0.05 & high-$z_s$ & $0.632^{+0.033}_{-0.040}$ & $0.929^{+0.738}_{-0.458}$ \\
0.10 & low-$z_s$ & $0.351^{+0.057}_{-0.037}$ & $5.142^{+2.685}_{-2.338}$ \\
0.10 & high-$z_s$ & $0.666^{+0.031}_{-0.038}$ & $1.102^{+0.929}_{-0.536}$ \\
0.15 & low-$z_s$ & $0.378^{+0.063}_{-0.051}$ & $5.981^{+2.478}_{-2.537}$ \\
0.15 & high-$z_s$ & $0.687^{+0.026}_{-0.032}$ & $1.352^{+1.253}_{-0.667}$ \\
0.20 & low-$z_s$ & $0.464^{+0.062}_{-0.076}$ & $6.982^{+2.084}_{-2.522}$ \\
0.20 & high-$z_s$ & $0.709^{+0.024}_{-0.028}$ & $1.062^{+1.108}_{-0.547}$ \\
\hline
\end{tabular}
\end{table}

\added{At the fiducial bandwidth $h=0.10$,} the resulting QSO slopes differ markedly. The low-$z_s$ bin gives
\begin{equation}
\gamma_{\rm QSO}=0.351^{+0.057}_{-0.037},
\end{equation}
whereas the high-$z_s$ bin gives
\begin{equation}
\gamma_{\rm QSO}=0.666^{+0.031}_{-0.038}.
\end{equation}
The corresponding curvature posteriors are
\begin{equation}
\Omega_K=5.142^{+2.685}_{-2.338}
\quad {\rm (low\ }z_s{\rm )},
\end{equation}
and
\begin{equation}
\Omega_K=1.102^{+0.929}_{-0.536}
\quad {\rm (high\ }z_s{\rm )},
\end{equation}
with central 95\% intervals $[1.088,9.579]$ and $[0.256,3.607]$, respectively. The low-$z_s$ QSO-slope posterior approaches the lower prior boundary $\gamma_{\rm QSO}=0.3$, while its curvature posterior has a long positive tail extending close to the upper prior boundary. We therefore do not interpret the difference as evidence for a specific physical evolution law. \added{Instead, the split indicates a lack of consistency of a single global QSO calibration within the adopted reconstruction and likelihood. Redshift-dependent QSO calibration, flux-selection effects, unresolved instrument/provenance mixtures, and kernel-reconstruction bias remain possible contributors; the present tests do not disentangle them.}

\begin{table}[!htbp]
\caption{Redshift-binned consistency test \added{using the full cluster covariance} for the nominal 100-pair sample \added{at the fiducial bandwidth $h=0.10$}.}
\label{tab:redshift}
\centering
\begin{tabular}{lccc}
\toprule
Bin & $N_{\rm pair}$ & $\gamma_{\rm QSO}$ (68\%) & $\Omega_K$ (68\%)\\
\midrule
Low $z_s$  & 50 & $0.351^{+0.057}_{-0.037}$ & $5.142^{+2.685}_{-2.338}$\\
High $z_s$ & 50 & $0.666^{+0.031}_{-0.038}$ & $1.102^{+0.929}_{-0.536}$\\
\bottomrule
\end{tabular}
\end{table}

\begin{figure}[!htbp]
\centering
\includegraphics[width=0.82\columnwidth]{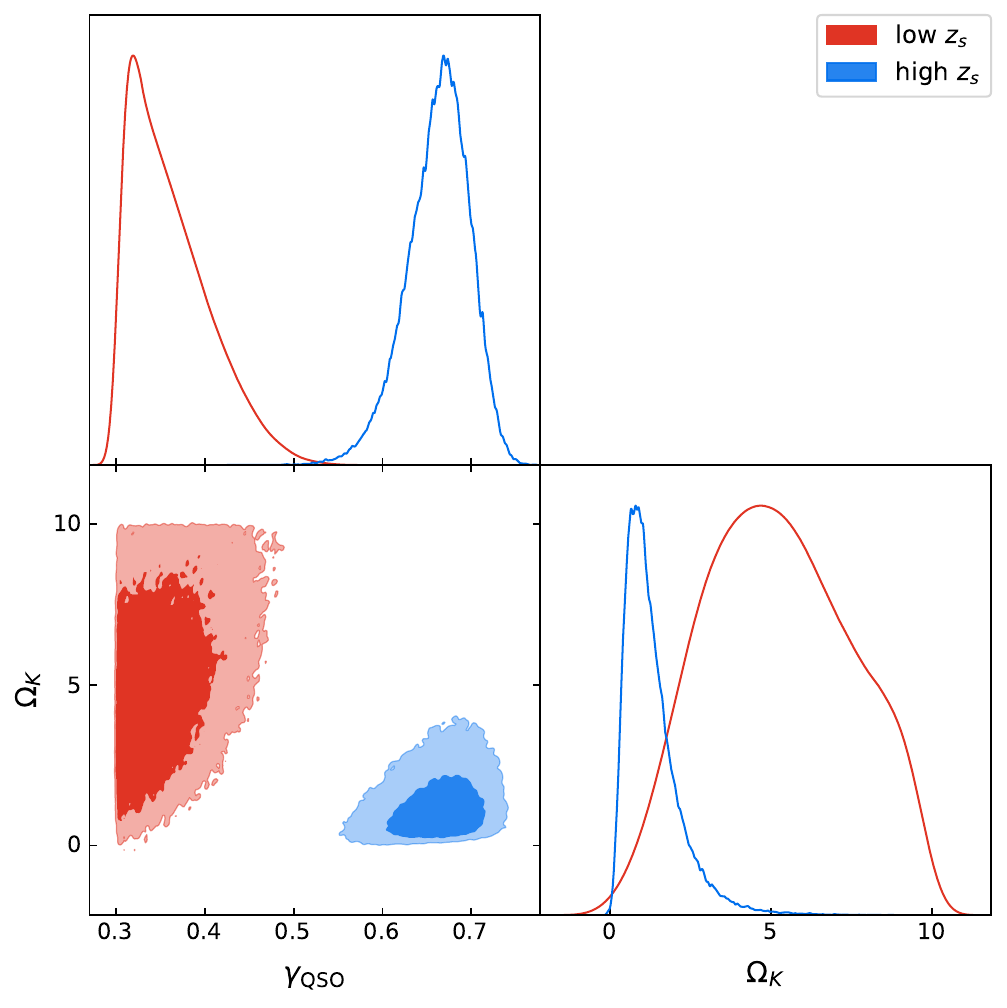}
\caption{Redshift-binned posterior comparison for the QSO slope and curvature. The two bins are separated at $z_{s,\rm med}=0.9335$ and each contains 50 matched lens systems. \added{The $h=0.10$ posteriors illustrate redshift-dependent calibration or reconstruction systematics, not a direct measurement of intrinsic QSO evolution.}}
\label{fig:redshift}
\end{figure}

\FloatBarrier
\subsection{Survey-motivated effective-sample \added{scaling illustration}}\label{sect:forecast}

The present machine-readable catalogs do not contain the selection functions and object-level survey/provenance labels needed to construct a literal Rubin/LSST or eROSITA mock catalog. \added{We therefore use an illustrative scaling of empirical posterior-width proxies rather than a survey-specific forecast. The exercise compares lens-rich and anchor-rich scenarios under an imposed scaling rule; it does not infer their precision from a future-data likelihood.} \added{These illustrative scalings assume unchanged statistical quality per effective object and hold the residual cross-probe systematics implicit in the present analysis fixed. They do not model additional calibration tensions or survey-selection effects in future data and must not be interpreted as formal Rubin/LSST or eROSITA uncertainty predictions.} 

For the nominal 5\% sample, the symmetrized 68\% curvature width is $\sigma_{\Omega_K}^{\rm joint}\simeq1.37$, whereas SGLS+QSO alone are effectively prior dominated with a symmetrized width $\sigma_{\Omega_K}^{\rm SGLS}\simeq6.72$. We define empirical \added{inverse-variance proxies}
\begin{equation}
F_{\rm SGLS}=\left(\sigma_{\Omega_K}^{\rm SGLS}\right)^{-2},\qquad
F_{\rm cl}=\left(\sigma_{\Omega_K}^{\rm joint}\right)^{-2}-F_{\rm SGLS},
\end{equation}
which give $F_{\rm SGLS}\simeq0.022$ and $F_{\rm cl}\simeq0.512$ for the current data. \added{The second quantity is an empirical increment associated with the absolute cluster-distance anchor within this likelihood, not an independent cosmological measurement. Because the SGLS+QSO width is prior dominated, these proxies are not likelihood Fisher information; fixed prior information does not increase with sample size. The extrapolation below is consequently a heuristic sensitivity illustration, not a statistically calibrated precision forecast.}

\added{For a purely illustrative scaling exercise, we use the empirical posterior-width proxies above as bookkeeping quantities and impose the following linear scaling rule:}
\begin{equation}
\sigma_{\Omega_K}^{-2}(N_{\rm pair},N_{\rm cl,eff})
=F_{\rm SGLS}\left(\frac{N_{\rm pair}}{100}\right)
+F_{\rm cl}\left(\frac{N_{\rm cl,eff}}{2.59}\right).
\label{eq:forecast_scaling}
\end{equation}
\added{The dependence on $N_{\rm pair}$ is imposed, not inferred from the prior-dominated SGLS+QSO result, and must not be interpreted as scaling fixed prior information with sample size. A quantitative prediction of the gain from additional lenses would require separating prior and likelihood information and modeling the latter.} This illustration compares a lens-rich Rubin/LSST-motivated direction with an anchor-rich direction relevant to larger X-ray cluster catalogs with independent SZ information. \added{Here $N_{\rm cl,eff}$ is an illustrative scaling coordinate normalized to the present prediction-only spectral dimension; it is not a measured Fisher-equivalent number of independent cluster constraints.}

\added{Under this imposed rule, Eq.~(\ref{eq:forecast_scaling}) gives only a small decrease in the illustrative width when $N_{\rm pair}$ grows at fixed $N_{\rm cl,eff}$:} increasing from 100 to 1000 effective matched systems at $N_{\rm cl,eff}=2.59$ changes $\sigma_{\Omega_K}$ from 1.37 to about 1.17. Increasing \added{the anchor scaling coordinate to 10} at the present lens sample size gives $\sigma_{\Omega_K}\simeq0.71$. A joint moderate scenario with $N_{\rm pair}=1000$ and $N_{\rm cl,eff}=25$ gives $\sigma_{\Omega_K}\simeq0.44$, while an ambitious $N_{\rm pair}=5000$, $N_{\rm cl,eff}=100$ scenario reaches about 0.22. Even these optimistic statistical scalings remain far less precise than CMB+BAO curvature constraints.

\begin{table}[!htbp]
\caption{Survey-motivated effective-sample \added{scaling illustration, assuming unchanged per-object statistical quality and no additional calibration or selection systematics. These values are neither formal Rubin/LSST or eROSITA uncertainty forecasts nor survey-yield predictions.}}
\label{tab:forecast}
\centering
\begin{tabular}{lccc}
\toprule
Scenario & $N_{\rm pair}$ & $N_{\rm cl,eff}$ & \added{Illustrative $\sigma_{\Omega_K}$}\\
\midrule
Current & 100 & 2.59 & 1.37\\
Lens-rich & 1000 & 2.59 & 1.17\\
Anchor-rich & 100 & 10 & 0.71\\
Joint moderate & 1000 & 25 & 0.44\\
Joint ambitious & 5000 & 100 & 0.22\\
\bottomrule
\end{tabular}
\end{table}

\begin{figure}[!htbp]
\centering
\includegraphics[width=0.86\columnwidth]{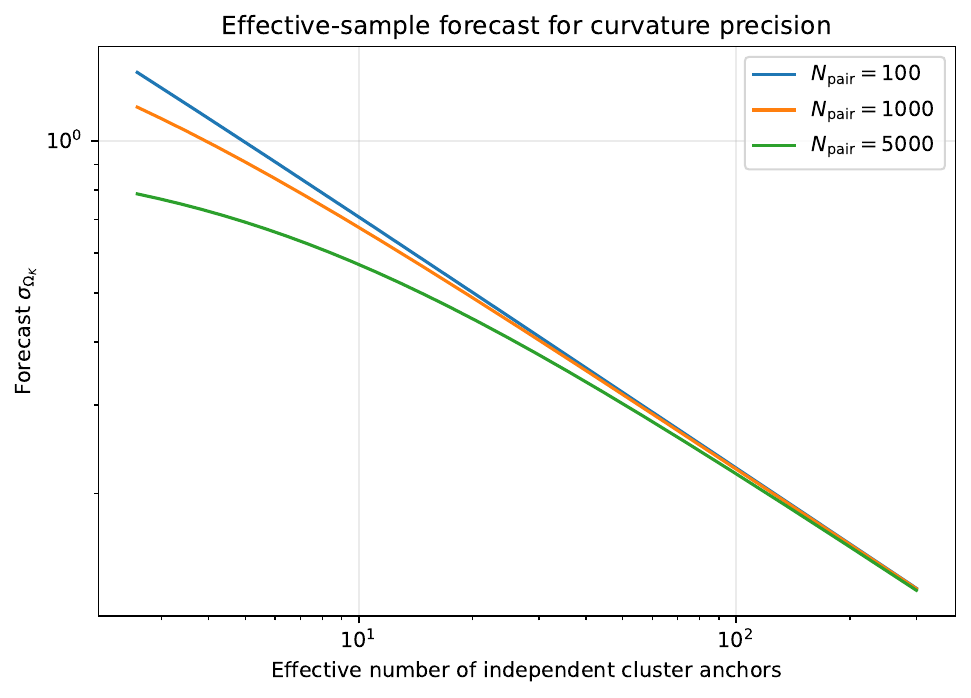}
\caption{\added{Illustrative curvature-width scaling as a function of the anchor coordinate $N_{\rm cl,eff}$, for three matched-lens sample sizes.} \added{The curves use the current SGLS+QSO-only and joint posterior widths as heuristic scaling proxies; they are neither likelihood Fisher forecasts nor formal survey uncertainty predictions.}}
\label{fig:forecast}
\end{figure}

\added{Within this imposed scaling, the anchor term dominates the illustrated improvement. The current SGLS+QSO posterior alone does not establish how its likelihood information will grow with sample size. These scenarios motivate improved absolute-distance anchors, QSO calibration, and cluster-modeling control, but quantifying future gains requires a survey-specific likelihood and systematic-error model.}

\subsection{Remaining data limitations}

Several relevant systematics cannot be evaluated robustly from the specific machine-readable parent tables used here, and we prefer to state those limitations rather than infer missing metadata. The adopted Risaliti--Lusso parent table does not carry the instrument/provenance labels required to reconstruct homogeneous SDSS--Chandra, XXL, and SDSS--4XMM matched subsamples. We therefore do not present an instrument-split posterior. Likewise, the adopted Bonamente table provides one angular-diameter-distance estimate per cluster, not a homogeneous alternative profile-by-profile distance set for all 38 systems; cluster triaxiality and non-isothermality are consequently treated as leading external systematics rather than replaced by an ad hoc profile correction.

A short-chain leave-one-out jackknife is not used as evidence in this analysis. Rather than quote a maximum excursion whose sampling error is comparable to the claimed effect, we base the robustness discussion on the deep-chain covariance, matching-threshold, \added{SGLS+QSO-only}, Bayesian-evidence, and redshift-binned controls reported above. The redshift split already demonstrates a strong calibration dependence, but the absence of object-level instrument/provenance labels prevents us from separating intrinsic redshift evolution from instrument- or selection-driven effects. An instrument-resolved reanalysis therefore remains a priority when the necessary metadata become available. \added{The importance of this distinction is supported by the sample- and redshift-dependent behavior of the UV/X-ray relation found by \citet{khadka22}; those findings motivate a provenance-resolved test but do not identify the cause of the split in our matched sample.}

\added{The covariance controls reported here concern the $38\times38$ cluster block. Shared QSO distance tracers can also correlate lensing and cluster residuals, and the cluster-block comparison does not separately quantify the impact of these cross-block terms. Thus ``full cluster covariance'' should not be read as a demonstration that every covariance in the joint residual vector has been validated. Robustness to this additional dependence requires a joint-residual or hierarchical analysis. The reported posterior intervals remain conditional on the adopted likelihood construction.}

\section{Conclusions}\label{sect:conclusion}

We have reconstructed and re-analyzed a model-independent late-time distance-consistency test based on strong gravitational lenses, UV/X-ray quasars, and SZ/X-ray galaxy-cluster distances. The main conclusions are:

\begin{enumerate}
\item Starting from 158 SGLS, the isothermal-reference condition $D_{A,ls}/D_{A,s}<1$ removes 30 geometrically inadmissible systems and leaves a fixed 128-system SGLS parent sample. Deterministic global matching to the 1598-QSO catalog then yields 100 independent SGLS--QSO pairs at the nominal 5\% distance threshold, without any $\gamma_{\rm QSO}$-dependent distance-order post-selection.

\item The full covariance among the 38 kernel-smoothed cluster predictions is essential. The prediction-only median off-diagonal correlation is 0.350 and the spectral effective dimension is 2.59. Ignoring the off-diagonal covariance shifts the curvature median from $\Omega_K=2.09$ to 5.21.

\item SGLS+QSO alone leave curvature essentially prior dominated, $\Omega_K=0.70^{+6.33}_{-7.12}$. The cluster distances provide most of the curvature constraining power, but they also make the result sensitive to cross-probe distance-scale consistency.

\item The nominal result \added{using the full cluster covariance} is $\Omega_K=2.085^{+1.634}_{-1.103}$, but the matching-threshold study shows a substantial 6\% assignment excursion. \added{Together with the strong covariance sensitivity and the dominant role of the cluster anchor, this demonstrates that the positive posterior offset is driven by cross-probe calibration tensions, sample-assignment effects, and the cluster anchor rather than by the intrinsic spatial geometry of the Universe.} We therefore do not interpret the positive posterior shift as evidence for an open Universe.

\item Nested sampling favors retaining the QSO slope as a free parameter over fixing $\gamma_{\rm QSO}=0.6$, with $\ln B_{\rm free/fixed}=4.50\pm0.17$. This preference is conditional on the present catalog construction and cross-probe likelihood.

\item A median-redshift split reveals a marked change in the QSO slope, from $\gamma_{\rm QSO}=0.351^{+0.057}_{-0.037}$ at low $z_s$ to $0.666^{+0.031}_{-0.038}$ at high $z_s$. \added{The separation persists over the tested bandwidth range $h=0.05$--$0.20$, although the inferred parameters remain sensitive to smoothing and estimator-induced redshift-dependent biases cannot be excluded.} Because the two fits share the same cluster anchor and the QSO table lacks instrument/provenance labels, we interpret this as a redshift-dependent calibration/systematics signal rather than a measurement of intrinsic QSO evolution.

\item \added{Under the imposed empirical scaling rule, increasing the matched-lens sample alone gives limited improvement at fixed anchor coordinate; this is not an inferred sample-size dependence of the prior-dominated SGLS+QSO likelihood.} Illustratively, $N_{\rm pair}=1000$ with $N_{\rm cl,eff}=25$ gives $\sigma_{\Omega_K}\simeq0.44$, while $N_{\rm pair}=5000$ and $N_{\rm cl,eff}=100$ gives about 0.22. \added{These are optimistic heuristic scalings with fixed residual systematics, not formal uncertainty predictions for Rubin/LSST or eROSITA.}

\item The Hubble-constant posterior is prior dominated and carries no independent information on the Hubble tension. \added{No meaningful constraint on the Hubble constant or the Hubble tension problem can be derived from this analysis.}
\end{enumerate}

The principal value of the present data set is therefore methodological: it exposes how strongly covariance, sample assignment, redshift-dependent QSO calibration, and QSO/cluster cross-calibration enter a nominally model-independent curvature inference. A future competitive application will require substantially larger lens samples, more independent absolute-distance anchors, and explicit control of QSO instrument/\allowbreak provenance and cluster sub-population systematics.

\section*{Data and code availability}
The observational catalogs used in this work are taken from the source publications cited in Sec.~\ref{sect:data}. The deterministic matching software used to construct the SGLS--QSO sample is publicly archived as \textsc{CosmoMatcher} \citep{cosmomatcher} at \url{https://doi.org/10.5281/zenodo.19695657} and is maintained at \url{https://github.com/HUJIAN0000/CosmoMatcher}.

\begin{acknowledgements}
We acknowledge the use of the GC compilation. This work is supported by Yunnan Youth Basic Research Projects 202501AT070439, the Local Universities Joint Special Project - General Project (Grant No. 202301A0070247) under the Basic Research Program, the National Natural Science Foundation of China (No.12473029), Dali Expert Workstation of Rainer Spurzem, Yunnan Academician Workstation of Wang Jingxiu (202005AF150025), China Manned Space Project (No. CMS-CSST-2021-A08), Guanghe Foundation (No. ghfund202407013470).
\end{acknowledgements}

\appendix

\section{Audit of the 158-system lens parent sample}\label{app:selection}

The 158-system table contains no missing entries in the numerical columns used in this work, no cases with $z_l\ge z_s$, and no duplicate names. The SGLS parent sample is defined by the single isothermal-reference geometric requirement $0<R_{\rm lens}(\gamma_{\rm lens}=2)<1$. Applying this cut removes 30 systems with $R_{\rm lens}\ge1$ and leaves exactly 128 systems. These 128 systems are then passed unchanged to the deterministic SGLS--QSO matching procedure described in Sec.~\ref{sect:matching}.

\section{Diagnostic of the removed QSO distance-order cut}\label{app:qso_cut}

As a separate diagnostic, an unconstrained matching run performed before the fixed 128-system SGLS parent definition was imposed produced 113 candidate matched systems. Evaluating the QSO distance ordering at $\gamma_{\rm QSO}=0.40,0.42,0.45,0.50,0.55,$ and 0.60 showed that 87 pairs are always ordered with $d_l<d_s$, 18 are always oppositely ordered, and 8 switch classification as $\gamma_{\rm QSO}$ changes. This diagnostic is not part of the final sample construction; it demonstrates why the adopted matching algorithm does not apply a hard QSO distance-order post-selection at a fixed value of the fitted slope.


\begin{thebibliography}{99}
\bibitem[Alam et al.(2017)]{alam17} Alam, S., et al. 2017, MNRAS, 470, 2617
\bibitem[Bernstein(2006)]{ber06} Bernstein, G. 2006, ApJ, 637, 598
\bibitem[Betoule et al.(2014)]{bet14} Betoule, M., et al. 2014, A\&A, 568, A22
\bibitem[Bonamente et al.(2006)]{bonamente06} Bonamente, M., Joy, M. K., LaRoque, S. J., Carlstrom, J. E., Reese, E. D., \& Dawson, K. S. 2006, ApJ, 647, 25
\bibitem[Cao et al.(2015)]{cao15} Cao, S., et al. 2015, ApJ, 806, 185
\bibitem[Cao et al.(2017)]{cao17} Cao, S., et al. 2017, A\&A, 606, A15
\bibitem[Clarkson et al.(2007)]{cla07} Clarkson, C., Cort\^es, M., \& Bassett, B. 2007, JCAP, 8, 011
\bibitem[Clarkson et al.(2008)]{cla08} Clarkson, C., Bassett, B., \& Lu, T. H.-C. 2008, Phys. Rev. Lett., 101, 011301
\bibitem[Coley et al.(2019)]{cole19} Coley, A. A., et al. 2019, JCAP, 05, 039
\bibitem[Collett et al.(2019)]{col19} \added{Collett, T., Montanari, F., \& R\"as\"anen, S. 2019, Phys. Rev. Lett., 123, 231101}
\bibitem[Cui et al.(2017)]{cui17} Cui, J. L., Li, H. L., \& Zhang, X. 2017, Sci. China Phys. Mech. Astron., 60, 080411
\bibitem[Di Valentino et al.(2020)]{di20} Di Valentino, E., Melchiorri, A., \& Silk, J. 2020, Nature Astronomy, 4, 196
\bibitem[Fan(1992)]{fan92} \added{Fan, J. 1992, J. Am. Stat. Assoc., 87, 998}
\bibitem[Foreman-Mackey et al.(2013)]{emcee13} Foreman-Mackey, D., Hogg, D. W., Lang, D., \& Goodman, J. 2013, PASP, 125, 306
\bibitem[Holanda et al.(2017)]{hol17} Holanda, R. F. L., Pereira, S. H., \& Jain, D. 2017, MNRAS, 471, 3079
\bibitem[Hu(2023)]{hu23a} Hu, J. 2023, ApJ, 948, 47
\bibitem[Hu et al.(2023)]{hu23b} Hu, J., et al. 2023, Phys. Rev. D, 108, 083024
\bibitem[Hu et al.(2026)]{hu26cddr} Hu, J., Liu, Y., Hu, J.-P., \& Li, Z. 2026, A\&A, 712, A17, doi:10.1051/0004-6361/202659255
\bibitem[Hu(2026)]{cosmomatcher} Hu, J. 2026, \textsc{CosmoMatcher}, Zenodo, doi:10.5281/zenodo.19695657
\bibitem[Khadka \& Ratra(2022)]{khadka22} Khadka, N., \& Ratra, B. 2022, MNRAS, 510, 2753
\bibitem[Leaf \& Melia(2018)]{leme18} Leaf, K., \& Melia, F. 2018, MNRAS, 478, 5104
\bibitem[Lenart et al.(2023)]{lenart23} Lenart, A. \L., Bargiacchi, G., Dainotti, M. G., Nagataki, S., \& Capozziello, S. 2023, ApJS, 264, 46
\bibitem[Li et al.(2018)]{li18} Li, Z. X., et al. 2018, ApJ, 854, 146
\bibitem[Liu et al.(2020)]{liu20} Liu, T. H., et al. 2020, MNRAS, 496, 708
\bibitem[Lusso et al.(2020)]{lusso20} Lusso, E., et al. 2020, A\&A, 642, A150
\bibitem[O'Colg\'ain et al.(2022)]{ocolgain22} O'Colg\'ain, E., Sheikh-Jabbari, M. M., Solomon, R., Bargiacchi, G., Capozziello, S., Dainotti, M. G., \& Stojkovic, D. 2022, Phys. Rev. D, 106, L041301
\bibitem[Park \& Ratra(2019)]{para19} Park, C. G., \& Ratra, B. 2019, ApSS, 364, 134
\bibitem[Planck Collaboration et al.(2016)]{planck16} Planck Collaboration, Ade, P. A. R., et al. 2016, A\&A, 594, A13
\bibitem[Planck Collaboration et al.(2020)]{planck20} Planck Collaboration, Aghanim, N., Akrami, Y., et al. 2020, A\&A, 641, A6
\bibitem[R\"as\"anen et al.(2015)]{ras15} R\"as\"anen, S., Bolejko, K., \& Finoguenov, A. 2015, Phys. Rev. Lett., 115, 101301
\bibitem[Riess et al.(2022)]{rie22} Riess, A. G., et al. 2022, ApJL, 934, L7
\bibitem[Risaliti \& Lusso(2019)]{ris19} Risaliti, G., \& Lusso, E. 2019, Nature Astronomy, 3, 272
\bibitem[Santos et al.(2017)]{san17} Santos, B., et al. 2017, JCAP, 02, 047
\bibitem[Shu et al.(2017)]{shu17} Shu, Y., et al. 2017, ApJ, 851, 48
\bibitem[Suzuki et al.(2012)]{suz12} Suzuki, N., et al. 2012, ApJ, 746, 85
\bibitem[Wang et al.(2020)]{wang20} Wang, B., et al. 2020, ApJ, 898, 100
\bibitem[Xia et al.(2017)]{xia17} Xia, J. Q., et al. 2017, ApJ, 834, 75
\bibitem[Yu \& Wang(2016)]{yuw16} Yu, H., \& Wang, F. Y. 2016, ApJ, 828, 85
\end{thebibliography}
\end{document}